# Fast dynamics in glass-forming salol investigated by dielectric spectroscopy

P. Lunkenheimer*, R. Wehn, M. Köhler, and A. Loidl

*Experimental Physics V, Center for Electronic Correlations and Magnetism, University of Augsburg, 86135 Augsburg, Germany*

A B S T R A C T

We analyze dielectric-loss spectra of glass forming salol extending up to 400 GHz allowing for the detection of the high-frequency minimum, where the fast critical dynamics predicted by the mode-coupling theory of the glass transition should prevail. Indeed, we find such a minimum which, moreover, well fulfills the critical scaling predicted by the theory. This includes the spectral shape of the minimum, the critical temperature dependence of the minimum frequency and amplitude, and the critical temperature dependence of the $\alpha$-relaxation rate at high temperatures. The minimum exponents $a$ and $b$ leading to a system parameter $\lambda \approx 0.63$ and the critical temperature $T_c = 256$ K are all in reasonable agreement with previous investigations of salol using different methods. Salol was one of the first materials where mode-coupling theory was tested and initial dielectric measurements were taken as an argument against the universal applicability of this theory.

## 1. Introduction

Salol is a typical example of a dipolar molecular glass former which can be easily supercooled. Investigations of such materials are commonly employed to help elucidating the mysteries of the glass transition and the glassy state of matter [1,2,3]. The mode coupling theory (MCT) is one of the most prominent theoretical approaches proposed to understand these mysteries [4]. One of its central predictions is the universal occurrence of fast critical dynamics, also termed fast $\beta$ process. It should arise in the time or frequency regime between the structural $\alpha$-relaxation process and the microscopic dynamics, observed in the vibrational frequency range and dominated by phonon-like or local molecular excitations. In susceptibility spectra, as obtained by experimental methods that are sensitive to the dynamics of the glass-forming entities (molecules in the case of salol), this dynamics should lead to a typical shallow minimum. Moreover, the temperature dependence and shape of this susceptibility minimum are predicted to obey characteristic scaling laws and to exhibit critical behavior revealing a critical temperature $T_c$ located above the glass temperature $T_g$. Within idealized MCT, $T_c$ marks a dynamic phase transition, which can be regarded as a kind of ideal glass transition below which the molecules are totally arrested [4]. However, in fact this transition is smeared out and less directly affecting the actual material properties due to additional hopping processes, treated in extended versions of the MCT [4,5]. Nevertheless, the critical behavior close to $T_c$ still should be observable in the experimental data.

Early tests of these far-reaching predictions were performed by inelastic scattering experiments like neutron [6,7] or light scattering [7,8,9,10,11], enabling to access the relevant frequency range in the GHz - THz range. Indeed, the existence of this minimum with significant self-similar enhancement above background contributions as well as the scaling and critical behavior were verified by these experiments. The MCT predicts that, for all experimental methods that couple to the glass-transition dynamics, this shallow susceptibility minimum should be present, exhibiting the same critical exponents and temperature. However, in the beginning stages of experimental tests of MCT, dielectric measurements did not detect the predicted minimum in the dielectric loss. Especially, such measurements performed for salol, extending into the 10 GHz range, were used to argue that there may be even no minimum at all in spectra of the dielectric loss and that there is no indication of critical behavior in these measurements

---

[12]. Thus, it seemed that the MCT predictions are not verified by dielectric spectroscopy which is the most applied experimental method to reveal information about glassy dynamics. However, due to the development of advanced coaxial reflection and transmission techniques and of submillimeter spectroscopy [13], it was later possible to measure the dielectric loss in an extended frequency range with high precision. Indeed, these experiments proved the validity of MCT by dielectric spectroscopy, at least for temperatures above $T_c$ [1,14,15,16,17,18]. The first such measurements were performed for the molecular glass formers glycerol [14] and for the ionic glass former [Ca(NO$_3$)$_2$]$_{0.4}$[KNO$_3$]$_{0.6}$ [15]. However, to our knowledge, until now no tests of MCT based on dielectric results were actually provided for salol. This is done in the present work, presenting dielectric data up to about 400 GHz, nicely revealing the predicted loss minimum and its scaling properties. Admittedly, the discussion about the dielectric high-frequency response of salol dates back more than 20 years [12,19]. Nevertheless, we think the present data are of interest to the community, especially because salol is an often investigated glass former and our results allow for a critical comparison with investigations by various other methods within this frequency/time range [10,20,21,22,23,24], also including relatively recent works [25,26]. We believe that the present results on salol are a significant further test of the predictive power even of the idealized MCT.

## 2. Experimental procedures

Salol (phenyl salicylate) with a purity of 99% was obtained from Aldrich Inc. Before the measurements, the crystalline material was heated above its melting point of about 41°C and filtered, which resulted in a negligible probability for devitrification. Broadband dielectric measurements of supercooled salol were performed using a combination of experimental methods [1,13,27]. In the high-frequency range beyond GHz, relevant for the present work, coaxial transmission measurements using a HP 8510 network analyzer were carried out, covering frequencies of 100 MHz - 25 GHz [13]. For this purpose, the sample material is filled into a specially designed coaxial line, sealed with Teflon discs. At frequencies 60 GHz < ν < 1.2 THz, a quasi-optical sub-millimeter spectrometer was employed [28], designed similar to a Mach - Zehnder interferometer. For temperature variation, closed-cycle refrigerators and N$_2$-gas cryostats were used.

## 3. Results and discussion

### 3.1. Broadband spectra

For an overview, Fig. 1 shows broadband loss spectra of salol, covering about 14 frequency decades [27,29]. As in the present work we are concentrating on the high-frequency response, in the following we only will briefly discuss the observations at frequencies below GHz. Under cooling, the dominating $\alpha$-relaxation peak strongly shifts to lower frequencies, which mirrors the glassy freezing of molecular motion when approaching the glass-transition temperature $T_g \approx 218$ K. Below about 233 K, a typical excess wing develops, well known to occur in various other glass formers. It shows up as a second power law at the high-frequency flank of the $\alpha$-relaxation peaks and was ascribed to a secondary relaxation process, whose loss peak is partly hidden under the dominating $\alpha$ peak [30,31,32,33,34]. The latter notion indeed is confirmed by the spectrum at 211 K, which was taken after an aging time of 6.5 days to ensure thermodynamic equilibrium. It exhibits a clear curvature, indicating a secondary relaxation process (see Ref. [32] for details).

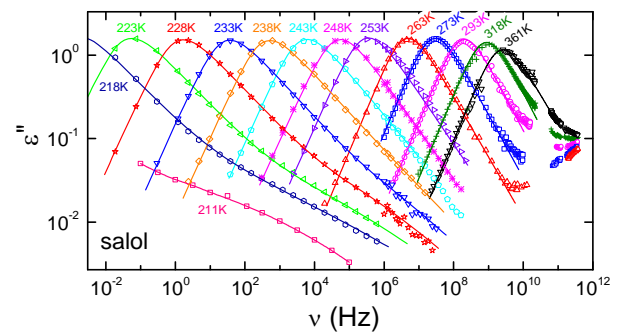

**Fig. 1.** Broadband dielectric loss spectra of glass forming salol at selected temperatures [27,29]. For $T \geq 243$ K the solid lines represent fits with a HN function, Eq. (1), ignoring the high-frequency minimum (for $T \geq 263$ K, $\alpha$ was zero, corresponding to a CD function). For smaller temperatures, the data were fitted by the sum of a HN function for the $\alpha$ relaxation and a CC function for the $\beta$ process. The line through the 211 K data is a guide to the eye.

The experimental data at low temperatures, $T \leq 238$ K, were fitted by the sum of a Havriliak-Negami (HN) function [35] for the $\alpha$ relaxation and a Cole-Cole (CC) [36] function for the $\beta$ relaxation, both commonly used empirical functions for the description of dielectric relaxation data (solid lines in Fig. 1). At higher temperatures, a single HN function was used for the fits. It is given by the formula

$$\varepsilon^* = \varepsilon_\infty + \frac{\varepsilon_s - \varepsilon_\infty}{[1 + (i\omega\tau)^{1-\alpha}]^\beta} \quad (1)$$

where $\varepsilon^* = \varepsilon' - i\varepsilon''$ is the complex permittivity, $\varepsilon_s$ is the static dielectric constant, $\varepsilon_\infty$ its high-frequency limit, and $\tau$ the relaxation time. $\alpha$ and $\beta$ are the spectral width parameters determining the symmetric and asymmetric broadening of the loss peaks, respectively. For $\beta = 1$, the formula is identical with the CC equation. For $T \geq 263$ K, satisfactory fits could be achieved with the parameter $\alpha$ fixed to zero. Then Eq. (1) corresponds to the often employed Cole-Davidson (CD) formula [37].

*3.2. Susceptibility minimum*

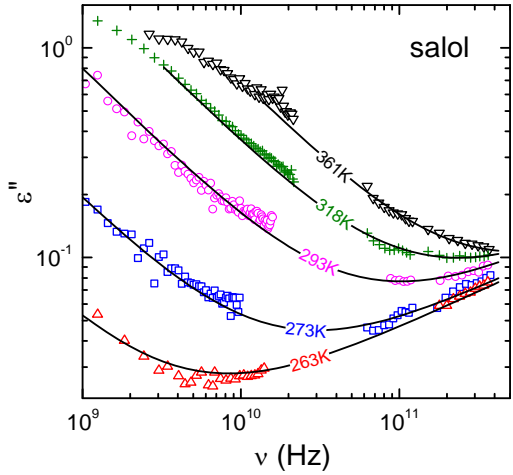

**Fig. 2.** Dielectric loss of salol in the frequency regime of the fast $\beta$ process [38]. The solid lines are fits with the mode-coupling theory, Eq. (2), with the same $a = 0.352$ and $b = 0.75$ for all temperatures, where $a$ and $b$ are linked by Eq. (3).

Figure 2 shows the evolution of the dielectric loss $\varepsilon''$ in the frequency regime between structural relaxation and the boson peak, expected to occur at about 1 THz [1], for temperatures between 263 and 361 K [38]. In this frequency regime, the dielectric loss of salol reveals a clear susceptibility minimum. According to MCT, it should approximately follow the sum of two power laws, namely:

$$\varepsilon'' = \frac{\varepsilon_{\min}}{a+b}\left[ a\left(\frac{\nu}{\nu_{\min}}\right)^{-b} + b\left(\frac{\nu}{\nu_{\min}}\right)^a \right] \quad (2)$$

Here the exponent $-b$ accounts for the increase towards the structural relaxation at low frequencies, and exponent $a$ for the increase at high frequencies. $\varepsilon_{\min}$ and $\nu_{\min}$ are the minimal dielectric loss and the minimum frequency, respectively. Indeed, good fits over up to 2.5 frequency decades can be achieved in this way (solid lines in Fig. 2). We find exponent parameters $a = 0.352$ and $b = 0.75$. Following the MCT prediction, both are directly linked to each other via the relation

$$\lambda = \frac{\Gamma^2(1-a)}{\Gamma(1-2a)} = \frac{\Gamma^2(1+b)}{\Gamma(1+2b)} \quad (3)$$

where $\lambda$ is the system parameter and $\Gamma$ denotes the Gamma function. In the present case, $\lambda \approx 0.63$. It should be noted that $a$ and $b$ are temperature independent, i.e., the minima at different temperatures are universal, and can be scaled onto each other.

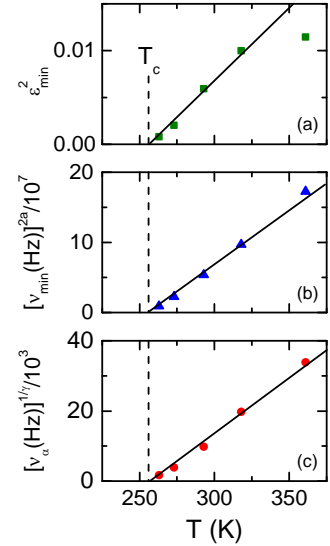

**Fig. 3.** Temperature dependence of the amplitude $\varepsilon_{\min}$ (a) and frequency $\nu_{\min}$ (b) of the $\varepsilon''(\nu)$ minimum and of the $\alpha$-relaxation rate $\nu_\alpha$ (c) of salol. The chosen representations should result in linear behavior according to the predictions of the MCT, Eqs. (4) - (6). The solid lines demonstrate a consistent description of all three quantities with a $T_c$ of 256 K.

On increasing temperature, the amplitude and frequency of the minimum both significantly increase. According to MCT, both quantities should follow critical behavior, namely

$$\nu_{\min} \propto (T - T_c)^{\frac{1}{2a}} \quad (4)$$

$$\varepsilon_{\min} \propto (T - T_c)^{1/2} \quad (5)$$

Notably, the critical exponent in Eq. (4) is directly related to one of the exponents determining the shape of the minimum. Figures 3(a) and (b) show an analysis of the present results in terms of the predicted MCT critical temperature dependence using representations that should result in linear behavior, crossing the abscissa at $T = T_c$. Obviously, both $\varepsilon_{min}$ and $\nu_{min}$ can be nicely described using the critical scaling predictions of MCT, consistently revealing a critical temperature of $T_c = 256$ K. In the pioneering work by Cummins and coworkers [10], by light scattering a system parameter $\lambda = 0.7$ as well as a critical temperature $T_c = 256$ K were determined, well consistent with the present results from dielectric spectroscopy. Comparable values were also obtained by various other investigations [21,22,23,24,25,39].

### 3.3. $\alpha$-relaxation parameters

According to MCT, the mean $\alpha$-relaxation time $\tau_\alpha$ or the corresponding relaxation rate $\nu_\alpha = 1/(2\pi\tau_\alpha)$ also should exhibit critical temperature dependence, namely

$$\nu_\alpha \propto (T - T_c)^\gamma \quad (6)$$

Here $\gamma$ is defined by $\gamma = 1/(2a) + 1/(2b)$, again directly relating the critical exponent to the shape of the loss minimum. Just as for the minimum parameters, Fig. 3(c) reveals that $\nu_\alpha(T)$ indeed is consistent with the predictions and with a critical temperature of 256 K as deduced from Figs. 3(a) and (b). Of course, one should be aware that this critical behavior is only expected to hold above $T_c$ and, consequently, in Fig. 3(c) only temperatures at $T > T_c$ are analyzed.

That Eq. (6) indeed is only valid at high temperatures is exemplified in Fig. 4. There we document the evolution of the relaxation time over the full temperature range [40] from the shortest time scales up to several 10000 s, where the equilibrium relaxation time was determined by aging the sample for 6.5 days [41]. Using $\tau_\alpha = 1/(2\pi\nu_\alpha)$, the solid line in Fig. 4 corresponds to the critical law (Eq. (6)) shown by the solid line in Fig. 3(c). Clearly, the divergence at $T_c$ predicted by idealized MCT is not documented in the data, but the dynamic phase transition is heavily smeared-out by hopping processes. The dashed line in Fig. 4 shows a fit of the experimental results with the empirical Vogel-Fulcher-Tammann (VFT) law, $\tau_\alpha \propto \exp[B/(T-T_{VF})]$, implying a divergence at the Vogel-Fulcher temperature $T_{VF} = 182$ K, significantly below $T_g \approx 218$ K [42,43,44]. As discussed in Ref. [40], for salol this often-employed function only roughly describes the overall behavior of $\tau_\alpha(T)$. Various alternative analyses of $\tau_\alpha(T)$ of salol were considered, e.g., in [27,29,40,45], but the discussion of these approaches is beyond the scope of the present work. Remarkably, in Ref. [46] it was shown that $\tau_\alpha(T)$ of salol in an extended temperature range below $T_c$ can be understood within a theoretical approach combining MCT and the random first-order transition theory, considering activated hopping processes.

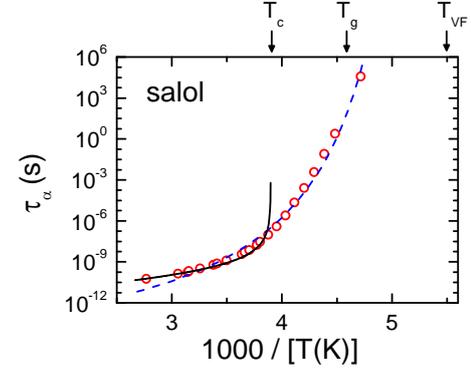

**Fig. 4.** Temperature dependence of the mean $\alpha$-relaxation time of salol shown in Arrhenius representation [40]. The solid line represents critical behavior as predicted by MCT and corresponds to the straight line in Fig. 3(c) (using $\tau_\alpha = 1/(2\pi\nu_\alpha)$). The dashed line is a fit with the VFT formula. The critical temperature of MCT, the glass-transition temperature, and the Vogel-Fulcher temperature are indicated by the arrows.

For completeness, in Fig. 5 we also provide the relaxation strength $\Delta\varepsilon = \varepsilon_s - \varepsilon_\infty$ and the width parameters $\alpha$ and $\beta$ of the $\alpha$ relaxation as resulting from the fits of the broadband spectra (Fig. 1). As commonly found in dipolar glass formers [1,2,47], $\Delta\varepsilon$ increases with decreasing temperature (Fig. 5(a)). Interestingly, the MCT also makes clear predictions concerning the temperature dependence of the relaxation strength: At $T > T_c$, the so-called non-ergodicity parameter $f$ is expected to be temperature independent. In the simplest case, for dielectric experiments $f$ may be assumed to be proportional to $\Delta\varepsilon$. In addition, for $T < T_c$ extended MCT predicts a square-root behavior, $f = c_1 + c_2 (T_c-T)^{1/2}$ ($c_1$ and $c_2$ are constants) [4]. The expected abrupt change of $f$ at $T_c$ from constant to critical behavior is often referred to as "cusp anomaly". Instead of $f \propto \Delta\varepsilon$, sometimes $f \sim \Delta\varepsilon T$ is assumed [48,49]. The latter relation includes a correction for the Curie law $\Delta\varepsilon \sim 1/T$ predicted by the Onsager theory, which is not taken into account by MCT. As shown by the solid line in the inset of Fig. 5, indeed this quantity is consistent with the cusp anomaly occurring at $T_c = 256$ K and with a constant behavior at $T > T_c$. However, one has to state that there are only few cases where the cusp anomaly has been verified in dielectric data (e.g., [49]). The reason may be that the temperature dependence of the relaxation strength is often dominated by dipolar correlations hampering an unambiguous comparison with theoretical predictions. By light scattering, the cusp anomaly of the non-ergodicity parameter of salol was verified in Refs. [20,39].

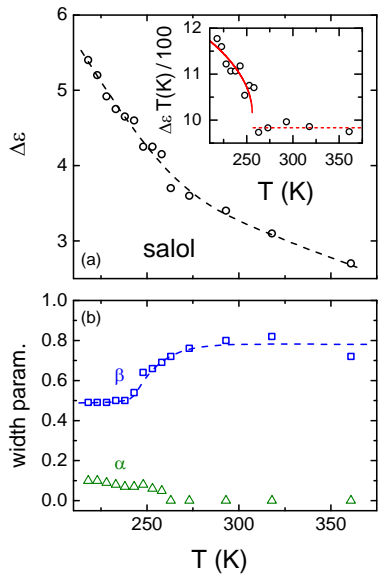

**Fig. 5.** Temperature dependence of the relaxation strength (a) and width parameters (b) of the $\alpha$ relaxation of salol. The lines in (a) and (b) are drawn to guide the eyes. The inset in (a) shows the relaxation strength multiplied by temperature. The solid line is a fit with the MCT square-root behavior (see text) with $T_c$ fixed to 256 K; the dashed line indicates approximately constant behavior above $T_c$.

The width parameter $\beta$ shown in Fig. 5(b) generally increases with temperature. The parameter $\alpha$ comes close to zero at high temperatures and, thus, was fixed to $\alpha = 0$ for temperatures above 260 K. Therefore, the data at these temperatures were effectively fitted with the CD formula (cf. discssion of Eq. (1)). MCT predicts a temperature-independent shape of the $\alpha$ relaxation peaks above $T_c$, implying so-called time-temperature superposition. The saturation of $\beta(T)$ documented in Fig. 5(b) indeed is compatible with this. Qualitatively similar behavior of salol was also found using other experimental methods [10,20,39]. High-temperature saturation of $\beta(T)$ was also reported by our group for various other glass formers (see, e.g., Refs. [1,47]).

One should note that in previous dielectric investigations of salol, partly quite different temperature dependences of the relaxation strength and width parameters were reported [12,48]. However, we want to remark that the inclusion of the $\beta$ relaxation at low temperatures in our analysis and the availability of experimental data at very high frequencies should enhance the precision of the obtained $\alpha$-relaxation parameters, compared to earlier work.

*3.4. Comparison with light scattering*

Finally, in Fig. 6 we compare our dielectric-loss spectra with the susceptibility spectra obtained by light scattering by Cummins and coworkers [10]. The latter, shown by the solid lines in Fig. 6, were scaled by a temperature-independent factor to approximately match the minimum amplitude. The relative temperature variation of this amplitude is very similar in both sets of data. At the lowest temperatures, the minimum frequencies of both methods also appear to be quite similar but with increasing temperatures deviations develop. Such deviations were also found in the few other molecular glass formers where such comparisons were made [1,38]. Moreover, in this scaled plot the $\alpha$-peak amplitudes of the scattering data are smaller. This implies different relative sensitivities of both experimental methods to the structural and fast dynamics. A stronger relative sensitivity of scattering methods to the fast process in the minimum region was also found for other molecular glass formers [1,38,49]. It can be understood by generalizations of MCT, incorporating orientational degrees of freedom [50,51].

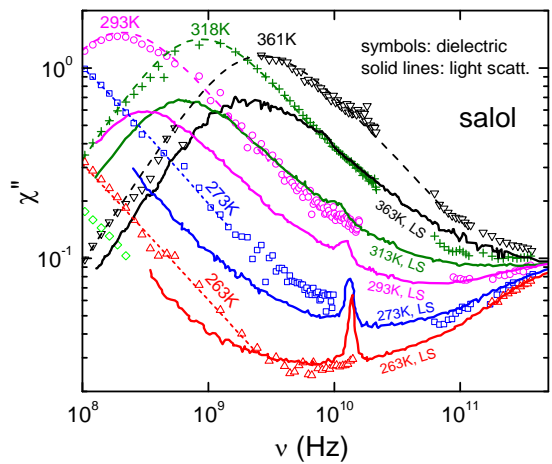

**Fig. 6.** Comparison of the present dielectric loss spectra of salol (symbols) with the susceptibility spectra deduced from light scattering (LS) as reported in [10] (solid lines) [29]. The latter were scaled by a temperature-independent factor to achieve a matching minimum amplitude for both data sets. The dashed lines are fits of the $\alpha$ peak with the CD equation (same as in Fig. 1). The sharp peak in the light-scattering data at about 15 GHz is an experimental artifact [10].

Overall, Fig. 6 reveals that the detected loss minima appear in a similar frequency range and have similar shape as for the light-scattering data. Especially, the left and right flanks have comparable slopes (i.e., exponents), just as expected by MCT relating these slopes to the probe-independent system parameter $\lambda$. The critical behavior of both data sets leads to an identical critical temperature of 256 K. We feel this is an excellent corroboration of the significance of the fast critical dynamics predicted by MCT.

## 4. Summary and Conclusions

In summary, our analysis of high-frequency dielectric spectra of glass-forming salol, extending well into the region of the predicted susceptibility minimum, reveals good agreement with the predictions of idealized MCT. Moreover, the properties of the $\alpha$ relaxation in this material also are consistent with the expected behavior. In fact, of all glass formers investigated by us up to the region of the loss minimum during the last 22 years [1,14,15,16,17,18,38,47], salol seems to be the one that is most consistent with the model predictions. The deduced MCT parameters $\lambda$ and $T_c$ are well compatible with previous findings obtained by different methods. We feel that these and corresponding results in other glass formers (e.g., [1,14,15,17,18,47]) document the predictive power of MCT. Lastly, we would like to remark that the present results finally settle the dispute on the existence of the high-frequency minimum and the critical dynamics in the dielectric response of salol [10,12,19].